\documentclass[twocolumn,showpacs,preprintnumbers,amsmath,amssymb,aps,a4paper]{revtex4}

\usepackage{amssymb,amsmath,amsbsy}
\usepackage{mathrsfs}
 \usepackage{graphicx}
 \usepackage{dcolumn}
 \usepackage{bm}
 \usepackage{verbatim}
 \usepackage{subfigure}
\usepackage{epstopdf}

 \newcommand{\fieldop}{a}

\newcommand{\dgg}{\dagger}

\begin{document}
  \title{Non-classical photon streams using rephased amplified spontaneous emission}

 \author{Patrick M. Ledingham, William R. Naylor, and Jevon
   J. Longdell} 
 \email{jevon.longdell@otago.ac.nz}
 \affiliation{Jack Dodd Centre for Photonics and Ultra-Cold Atoms, Department of Physics, University of Otago, Dunedin, New Zealand.}
\author{Sarah E. Beavan, and Matthew J. Sellars}
\affiliation{Laser Physics Centre, RSPhysSE, Australian National University, Canberra, ACT 0200, Australia}
\date{\today}
\begin{abstract}
  We present a fully quantum mechanical treatment of optically
  rephased photon echoes. These echoes exhibit noise due to
  amplified spontaneous emission, however this noise can be seen as a
  consequence of the entanglement between the atoms and the output
  light. With a rephasing pulse one can get an ``echo'' of the
  amplified spontaneous emission, leading to light with nonclassical
  correlations at points separated in time, which is of interest in
  the context of building wide bandwith quantum repeaters. We also
  suggest a wideband version of DLCZ protocol based on the same ideas.
\end{abstract}

\maketitle

\section{\bf Introduction\label{sec:Intro}}

In order to extend the range of quantum key distribution, quantum
networks and tests of Bell inequalities, a method for efficiently
generating entanglement over large distances is required. To achieve
this goal a quantum repeater is necessary \cite{brie98}. Such
repeaters are generally based on methods for entangling one light
field entangled with another at a later point in time. This has led to
increasing interest in quantum memories for light, which in
conjunction with pair sources would achieve this. Many impressive
experiments have been performed in the area of quantum memories and
repeaters. The quantum state of a light field has been stored in a
vapour cell with high fidelity, and then measured at a later time
\cite{juls04}. Single photons and squeezed states have been stored and
recalled \cite{eisa05,chan05,appe08}, and non-classical interference
of the light from distant ensembles has been observed
\cite{chan07}. Entanglement \cite{moeh07} of and
teleportation\cite{olms09} between two distant trapped ions has been
achieved using an optical channel, using DLCZ type measurement induced
entaglement.

Photon echoes have a long history of use in classical signal processing with
light \cite{kurn64,moss82}. There are now a number of proposals and
experiments \cite{mois01,hete08,alex07,stau07,reid08,hoss09} related to the
development of photon echo based quantum memories. A distinct
advantage of echo based techniques is that they are multimode
\cite{simo07}.

Current photon echo quantum memory techniques, all involve some
modification of the inhomogeneous broadening profile. This imposes
limits on 
the range of suitable materials. It would be much more convenient to
use something akin to the standard two pulse echo as a quantum
memory which doesn't require such modification. 

The paper is arranged in three sections. In Sec. \ref{sec:QMBE} we
present the quantum mechanical Maxwell-Bloch equations for atomic and
photonic fields. Then in Sec. \ref{sec:2PE}, as an example of this
formalism, we present an analysis of the standard two pulse photon
echo and its applicability as a quantum memory. The two pulse echo as
a quantum memory has already been investigated by others
\cite{rugg08}. We revisit the problem here and show that this protocol
fails as a quantum memory due to the strong rephasing pulse, which
inverts the medium and causes additional noise on the output photonic
fields. Finally in Sec. \ref{sec:RASE}, after exploring the origin of
this noise, we propose that this noise can be rephased to lead to time
separated, temporally multimode, wide bandwith photon streams with
non-classical correlations. So while a standard two pulse echo fails
as a quantum memory, rephased amplified spontaneous emission (RASE)
can be utilized in the DLCZ protocol \cite{dlcz}. With this modified
DLCZ protocol, the inhomogeneous broadening no longer limits the time
between the write and read pulses but instead increases the bandwidth
of the process. This is of significance to current experiments, were
the inhomogeneous broadening is an
issue\cite{feli05,laur07,kuzm03,chou05,reid06}.

\section{\bf Quantized Maxwell-Bloch Equations\label{sec:QMBE}}

We shall model an inhomogeneously broadened collection of two level
atoms interacting with a 1-D field propagating in one direction, with
the following quantum Maxwell-Bloch equations:
\begin{align}
	\frac{\partial}{\partial t} \hat{\sigma}_-(z,\Delta,t)  =& i \Delta \,\hat{\sigma}_-(z,\Delta,t) - i\,\hat{\fieldop}(z,t)\, \hat{\sigma}_z(z,\Delta,t) \label{eq:mb1}\\
	\frac{\partial}{\partial t} \hat{\sigma}_z(z,\Delta,t)  =& i\,\hat{\fieldop}(z,t)\, \hat{\sigma}_-(z,\Delta,t) \notag \\
	& -i\,\hat{\fieldop}^\dagger(z,t)\, \hat{\sigma}_+(z,\Delta,t) \label{eq:mb2}\\
	 \frac{\partial}{\partial z} \hat{\fieldop}(z,t)  =& \frac{i \alpha}{2 \pi} \int_{-\infty}^{\infty} \hat{\sigma}_-(z,\Delta,t) \,d\Delta,\label{eq:mb3}
\end{align}
where $\hat{\sigma}_{+,-,z}$ represent the quantum atomic spin
operators, $\hat{\fieldop}$ is the quantum optical field operator,
$\alpha$ is the optical depth parameter, which depends on the coupling
between the atoms and the field and on the
atom density. The parameter $\Delta$ is the detuning from some chosen
resonant frequency and $z$ is the distance along the propagation direction.
The operators have the following commutation relations:

\begin{eqnarray}
{[\hat\fieldop(z,t),\hat\fieldop^\dagger(z,t')]} &=& \delta(t-t')\\
{[\hat\sigma_i(z,\Delta,t),\hat\sigma_j(z',\Delta',t)]} &=& \frac{2\pi}{\alpha}\epsilon_{ijk}\hat\sigma_k(z,\Delta,t)\notag \\
\times \delta(z-z')\delta(\Delta-\Delta')
\end{eqnarray}
As can be seen from  Eq.~\ref{eq:mb3}, we take the density of atoms as
a function of frequency to be a constant. In the case of rare earth
ion dopants, the inhomogeneous broadening can be many times larger than
the homogeneous linewidths, and as a result in most experiments
without holeburnt features this is a good approximation.

The above Maxwell-Bloch equations can be derived by dividing the
atomic ensemble into thin slices and then modelling each slice as a
small collection of atoms inside a Fabry-Perot cavity, using standard
input-output theory \cite{gard85}. Taking the limit as
reflectivity of the mirrors go to zero one arrives at
Eq.~(\ref{eq:mb1}--\ref{eq:mb3}), where $\hat{\fieldop}(z,t)$ is the input
field at the left hand side of the cavity and $\hat{\fieldop}(z+d\,z,t)$ is
the output field at the right hand side of the cavity.

\begin{figure}
\includegraphics[width=0.5\textwidth]{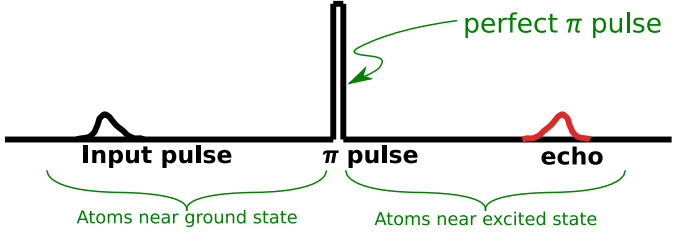}
\caption{(Color online) \label{fig:twopulseapprox}A two pulse photon echo sequence showing the approximations
  made in the treatment, a weak first pulse is
  applied to the system and is recalled using and ideal $\pi$-pulse.}
\end{figure}

\section{\bf The Two Pulse Photon Echo\label{sec:2PE}}

The first application of our quantum Maxwell-Bloch equations will be
in analysing a memory based on a two pulse photon echo. The
Maxwell-Bloch equations are non-linear and in general difficult to
solve analytically, however following work done with the semiclassical
Maxwell-Bloch equations \cite{tsan03} one can make reasonable approximations
that simplify the situation greatly. First we shall assume the input
pulse is weak and is much smaller than a $\pi$ pulse.
In this case all the atoms will stay near their ground state
($\sigma_z \approx -1$) and we can approximate the 
atomic lowering operator $\sigma_{-}$ as a harmonic oscillator
field $D_g$. The result are linear equations which we shall refer to
as the ground state Maxwell-Bloch equations,

\begin{eqnarray}
\label{equ:gs1}
	\frac{\partial}{\partial t} \hat{D}_{g}(z,\Delta,t) & =& i \Delta \,\hat D_{g} (z,\Delta,t) +i\, \hat{\fieldop}(z,t) \\
\label{equ:gs2}
	 \frac{\partial}{\partial z} \hat{\fieldop}(z,t) & =& \frac{i \alpha}{2 \pi} \int_{-\infty}^{\infty} \hat D_{g} (z,\Delta,t) \,d\Delta .
\end{eqnarray}

Eq.~\ref{equ:gs1} is just a first order linear equation with solution,
\begin{align}
	\hat{D}_g(z,t,\Delta) &= -i \int_{-\infty}^{t} dt' \; \notag 
        \hat{\fieldop}(z,t')
  e^{i \Delta(t-t')} \\   
	 &  \qquad + e^{i \Delta t} \hat{D}_{g0}(z,\Delta) \;,\label{eq:gs_soln1}
\end{align}
where $\hat{D}_{g0}(z,\Delta) $ is an initial condition. Taking the Fourier transform of Eqs.~\ref{equ:gs2} and \ref{eq:gs_soln1}  and substituting, one arrives at the following expression:
\begin{align}
	\frac{\partial}{\partial z} \hat{\fieldop}(z,\omega) &= \frac{-\alpha}{2\pi} \int_{-\infty}^{\infty} d\Delta \;  \hat{\fieldop}(z,\omega) \bigg[ \frac{1}{i(\omega - \Delta)} + \pi \delta(\omega - \Delta) \bigg] \notag \\
	& \qquad + \frac{i \alpha}{\sqrt{2\pi}} \int_{-\infty}^{\infty}  d\Delta \; \delta(\omega - \Delta) \hat{D}_{g0}(z,\Delta) \notag \; , \\
	&= \frac{-\alpha}{2}\hat{\fieldop}(z,\omega) + \frac{i \alpha}{\sqrt{2\pi}} \hat{D}_{g0}(z,\omega),  \; \label{equ:gsFT}
\end{align}
where $ \delta(\omega)$ is the Dirac delta function. Solving Eq. \ref{equ:gsFT} and fourier transforming back to the time domain we get
\begin{align}
\hat{\fieldop}(z,t) &= \hat{\fieldop}(0,t) e^{-\alpha z/2}  \notag \\
& \qquad  + \frac{i \alpha}{\sqrt{2\pi}} \int_0^z dz' \; e^{\alpha(z' - z)/2} \; \hat{D}_{g0}(z',t) \; ,\label{eq:gs_soln2}\notag \\
\end{align}
where $\hat{\fieldop}(0,t)$ denotes the input photonic field. Eqs.~\ref{eq:gs_soln1} and \ref{eq:gs_soln2} form the ground state solutions for all input times.

After the $\pi$-pulse the atoms are all very close to the excited
state ($\sigma_z \approx +1$) in which case we can approximate
$\sigma_{+}$ by a harmonic oscillator field $D_e$. This gives us the
excited state Maxwell-Bloch equations.

\begin{eqnarray}
	\label{equ:ex1}
	\frac{\partial}{\partial t} \hat{D}_{e}^\dagger(z,\Delta,t) & =& i \Delta \,\hat D_{e}^\dagger (z,\Delta,t) -i\, \hat{\fieldop}(z,t) \\
	\label{equ:ex2}
	 \frac{\partial}{\partial z} \hat{\fieldop}(z,t) & =& \frac{i \alpha}{2 \pi} \int_{-\infty}^{\infty} \hat D_{e}^\dagger (z,\Delta,t) \,d\Delta .
\end{eqnarray}

We treat the $\pi$ pulse as being a perfect $\pi$ leading to the transformation $\hat D_e \leftarrow \hat D_g$. We will discuss the treatment of the perfect $\pi$ pulse later in the text.

Bringing Eqs.~\ref{equ:ex1} and \ref{equ:ex2} through the same mathematical process as Eqs.~\ref{equ:gs1} and \ref{equ:gs2}, we arrive at the excited state solutions:  

\begin{align}
	\hat{D}_e^\dagger(z,t,\Delta) &= i \int_{-\infty}^{t} dt' \; 
\hat{\fieldop}(z,t')
e^{i \Delta(t-t')} 
	 + e^{i \Delta t} \hat{D}_{e0}^\dagger(z,\Delta) \;,\label{eq:es_soln1}
\end{align}
\begin{align}
\hat{\fieldop}(z,t) &= \hat{\fieldop}(0,t) e^{\alpha z/2}  \notag \\
& \qquad  + \frac{i \alpha}{\sqrt{2\pi}} \int_0^z dz' \; e^{\alpha(z -  z')/2} \; \hat{D}_{e0}^\dagger(z',t) \; ,\label{eq:es_soln2}
\end{align}
where $\hat{\fieldop}(0,t)$ and $ \hat{D}^\dagger_{e0}(z,\Delta) $ are initial conditions for the photonic and atomic excited fields respectively.

Matching the ground (Eqs.~\ref{eq:gs_soln1} and~\ref{eq:gs_soln2}) and
excited (Eqs.~\ref{eq:es_soln1} and~\ref{eq:es_soln2}) state solutions
at the point the $\pi$-pulse is applied we get a complete
solution. The efficiency is $\sinh^2(\frac{\alpha z}{2})$ and in the
limit of large optical depths high efficiencies are
possible. Physically, this is because the photon echo is produced in the first
piece of the sample and then gets amplified as it propagates through the
inverted medium. The noise on the output can be quantified by
considering the case of no input pulse, then the output will be
amplified spontaneous emission (ASE), simply
the vacuum noise amplified by the gain of $\exp(\alpha z)$ of the
inverted ensemble. In the case of no input pulse, we get an incoherent
output field with $\langle \fieldop^\dagger(t)\fieldop(t') \rangle
=\delta(t-t')(\exp(\alpha l)-1)$.

It is interesting to consider the source of this noise. In the model
we have no dissipation and so the total system evolves through pure
states. 

Eq.~\ref{eq:gs_soln1} is analogous the output of a beam splitter.
The input fields being light and atoms, with the output fields
consisting of combinations of photonic and atomic excitations. One can
see that the addition of atomic excitations in the solution is
necessary for the conservation of the commutation relations due to the
input photonic field decaying away at large $\alpha l$.  The excited
state solution is analogous to a non-degenerate parametric amplifier
\cite{wallsnmilburn}, here the input field is amplified, and the
comutation relations are preserved by the addition of atomic creation
operators.  The state of one output mode is only mixed if the other is
traced over, if the system is viewed as a whole one has an entangled
state. In the next section we show that by applying a rephasing pulse to the ensemble
we can turn the excitation of the atoms back into light, leading to
streams of photons with highly non-classical correlations between two
points separated in time.
\section{\bf Rephased Amplified Spontaneous Emisson\label{sec:RASE}}
\begin{figure}
\centering\includegraphics[width=0.5\textwidth]{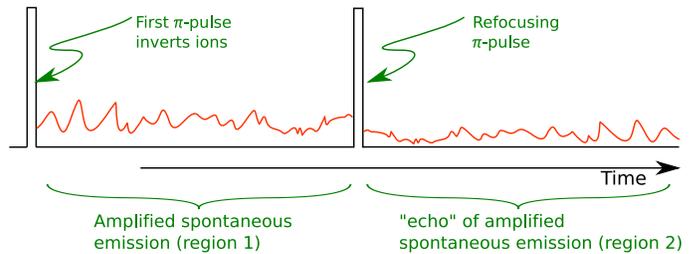}
\caption{(Color online) Two $\pi$ pulse photon echo sequence proposed
  for generating rephased amplified spontaneous emission (RASE).}
\label{fig:2pipulse}
\end{figure}
Now we consider the two $\pi$ pulse sequence shown in
Fig.~\ref{fig:2pipulse}. For region~1 the atoms will be inverted due
to the first $\pi$ pulse and hence Eqs.~(\ref{equ:ex1},
\ref{equ:ex2}) will apply. For region~2 the atoms will be near the
ground state due to the refocusing $\pi$ pulse, hence Eqs.~(\ref{equ:gs1}, \ref{equ:gs2}) describe the dynamics. We take the
second $\pi$-pulse to occur at $t=0$.

The solution for the light in region~1 is given by
Eqs.~\ref{eq:gs_soln1} and~\ref{eq:gs_soln2} and the solution in
region~2 is given by Eqs.~\ref{eq:es_soln1}
and~\ref{eq:es_soln2}. For boundary conditions we take the incident
field, $\hat{\fieldop}(0,t)$, to be in it's vacuum state as we do for
the initial condition $D_{e0}(z,\Delta)$. The initial condition for
region~2 we get from the final condition for region~1:
\begin{align}
  \label{eq:reg2_ic}
	\hat{D}_{g0}(z,\Delta) &= i e^{\alpha z/2} \int_{-\infty}^{0} dt' \; \hat{\fieldop}^\dgg(0,t')  \, e^{i \Delta t'}  \notag \\
	&+ \frac{\alpha}{\sqrt{2\pi}} \int_{-\infty}^{0} dt' \;  \, e^{i \Delta t'} \int_0^z dz' \; e^{\alpha(z - z')/2} \; \hat{D}_{e0}(z',t') \notag \\
	&+ \hat{D}_{e0}(z,\Delta) \; .
\end{align}

These boundary and initial conditions substituted in
Eqs.~(\ref{eq:gs_soln1}--\ref{eq:es_soln2}) give a complete
analytic solution of the linearised Maxwell-Bloch equations.

To show that the photon streams described by these solutions have
non-classical correlations we consider,

\begin{equation}
	R \equiv \frac{p(t_1,t_2)^2}{p(t_1,t_1) \, p(t_2,t_2)} ,
\end{equation}
where $	p(t_i,t_j) = \langle \, \hat{\fieldop}^\dgg(l,t_i) \,
\hat{\fieldop}(l,t_i) \, \hat{\fieldop}^\dgg(l,t_j) \, \hat{\fieldop}(l,t_j)
\, \rangle$. For classical fields the Cauchy-Schwartz inequality states that
$R\le1$ \cite{wallsnmilburn}. Considering times equally separated
about the second $\pi$-pulse, from the expression for the output fields derived above we get 
\begin{equation}
	R(\alpha l) = \Bigg[\frac{1}{2} + \frac{\alpha l \, + \cosh \!\! \big(\alpha l\big)}{4 \, \sinh \!\! \Big(\frac{\alpha l}{2}\Big) \big[e^{\alpha l} - 1 \big]}\Bigg]^2
\end{equation}
\begin{figure}
\includegraphics[width=0.3\textwidth]{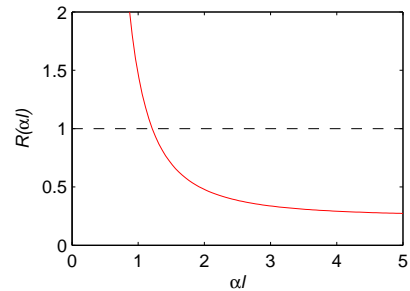}
\caption{(Color online) Plot of $R(\alpha l)$, show the violations of the Cauchy-Schwartz
inequality for small optical depths.}
\label{fig:Rofz}
\end{figure}

Figure \ref{fig:Rofz} shows that for small optical depths $(\alpha l <
1)$ the output at times equally separated from the refocusing
$\pi$-pulse has nonclassical correlations. It should be pointed out that the entanglement
between times is not perfect because the echo efficiency is not
$100\%$. The detection of a photon in Region 2 at a particular time
means that there must have been one in Region 1 at the matching time,
however the converse is not true.

An advantage that  RASE has is it's potential implementation in a larger range of systems. This is in contrast with the implementation of current photon echo quantum memories.  Current quantum memory echo techniques use very fine spectral features prepared in the inhomogeneous line, rather than the natural inhomogeneous profile and optical rephasing pulses. Indeed the fact that AFC\cite{reid08} and CRIB\cite{mois01,hete08} type echoes ideally want homogeneously broadened ensembles, has led to their investigation in non-solid-state systems \cite{hete08e}.  When selecting a rare earth ion system, one finds that the systems long lived spectral holes, such as europium or praseodymium, have inconvenient wavelengths ($\approx 580$~nm and $\approx 606$~nm), in systems which are much more compatible with optical fibers and diode lasers the holes are much more transient, making high fidelity operation difficult at best.

\section{\bf Imperfect $\pi$ pulses\label{subsec:some}}
So far we have treated the $\pi$ pulses as ideal and the effect of
non-ideal $\pi$ pulses needs to be considered. It is feasible
to make a pulse such that afterward one can make the approximation
$\sigma_z\approx1$ especially as we are interested in optically thin
samples. The ability to do this in optically thick samples is also
helped by the area theorem, which states that a $\pi$ pulse remains a
$\pi$ pulse as it propagates through a medium \cite{hahn71,rugg08}. In
the situation where $\sigma_z\approx1$ after the pulse, we can model
our non-ideal $\pi$ pulse as the combination of an ideal $\pi$ pulse
and some excitation of the $\hat D_e$ field. This excitation of the
$\hat D_e$ field will be temporally brief, and if the inhomogeneous
broadening is flat the ensemble of atoms will quickly dephase leading
to no net polarization in the ensemble shortly after the $\pi$
pulse. This means that the excitation produced by the imperfect $\pi$
pulse will no longer interact with the optical field (unless it is
rephased by another strong pulse). The ability to prepare an ideal
inverted medium for classical information processing has been
investigated experimentally \cite{zafa07}.

\section{\bf Phase Matching}

%
%
The treatment so far has involved only one spatial dimension. One way to consider the effect of phase matching is by extending to a 3D treatment in the paraxial approximation. In this case we replace $a(z,t)\rightarrow a(z,\mathbf{k}_t,t)$ and $\sigma_-(z,\Delta,t) \rightarrow \sigma_-(z,\mathbf{\rho},\Delta,t)$ where $\mathbf{k}_t = (k_x,k_y)$ is the transverse wavevector and $\mathbf{\rho}=(x,y)$ is the transverse position. Our linearized Maxwell Bloch equations for the ground state become

\begin{eqnarray}
\label{equ:gs1parax}
	\frac{\partial}{\partial t}
        \hat{D}_{g}(z,\mathbf{\rho},\Delta,t) & =& i \Delta \,\hat
        D_{g} (z,\mathbf{\rho},\Delta,t)\nonumber\\
&& +\frac{i}{4\pi^2} \int d^2\mathbf{k_t}\,
\hat{\fieldop}(z,\mathbf{k}_t,t)e^{i \mathbf{k}_t.\mathbf{\rho}}
\\
\label{equ:gs2parax}
	 \frac{\partial}{\partial z} \hat{\fieldop}(z,\mathbf{k}_t,t)
         & =& \nonumber\\
\frac{i \alpha}{2 \pi} \int d^2\mathbf{\rho}\int_{-\infty}^{\infty}&
d\Delta&\,\hat D_{g} (z,\mathbf{\rho},\Delta,t)e^{-i\mathbf{k}_t.\mathbf{\rho}}.
\end{eqnarray}

Fourier transforming the atomic operators along the transverse
dimensions by defining 
\begin{equation}
\hat {D}_{g}(z,\mathbf{k}_t,\Delta,t) = \int d^2\mathbf{\rho}\, \hat{D}_{g}(z,\rho,\Delta,t) \exp(-i \mathbf{k}_t.\rho)
\end{equation}
leads to Maxwell Bloch equations which are diagonal in the transverse
wave vector

\begin{eqnarray}
\label{equ:gs1parax_diag}
	\frac{\partial}{\partial t}
        \hat{D}_{g}(z,\mathbf{k}_t,\Delta,t) & =& i \Delta \,\hat
        D_{g} (z,\mathbf{k}_t,\Delta,t)
+i \hat{\fieldop}(z,\mathbf{k}_t,t)
\\
\label{equ:gs2parax_diag}
	 \frac{\partial}{\partial z} \hat{\fieldop}(z,\mathbf{k}_t,t)
         & =& 
\frac{i \alpha}{2 \pi} \int_{-\infty}^{\infty}
d\Delta\,\hat D_{g} (z,\mathbf{k}_t\Delta,t).
\end{eqnarray}
For the excited state Maxwell Bloch, the same procedure gives
\begin{eqnarray}
	\label{equ:ex1parax}
	\frac{\partial}{\partial t} \hat{D}_{e}^\dagger(z,\mathbf{k}_t,\Delta,t) & =& i \Delta \,\hat D_{e}^\dagger (z,\mathbf{k}_t,\Delta,t) -i\, \hat{\fieldop}(z,-\mathbf{k}_t,t) \\
	\label{equ:ex2parax}
	 \frac{\partial}{\partial z} \hat{\fieldop}(z,\mathbf{k}_t,t) & =& \frac{i \alpha}{2 \pi} \int_{-\infty}^{\infty} \hat D_{e}^\dagger (z,-\mathbf{k}_t,\Delta,t) \,d\Delta .
\end{eqnarray}

In the situation where the $\pi$-pulse is applied off axis the phase of the $\pi$-pulse depends on the transverse position leading to the 
transformation 

\begin{equation}
\hat D_e(z,\mathbf{\rho},\Delta,t) \leftarrow \hat
D_g(z,\mathbf{\rho},\Delta,t)\exp(2i\mathbf{k}_\pi.\mathbf{\rho}),
\end{equation}

or after fourier transforming

\begin{equation}
\label{equ:paraxpi}
\hat D_e(z,\mathbf{k}_t,\Delta,t)  \leftarrow \hat
D_g(z,\mathbf{k}_t-2\mathbf{k}_\pi,\Delta,t).
\end{equation}

With the RASE pulse sequence described in Fig.~\ref{fig:2pipulse}, the
phase of the first $\pi$ pulse doesn't matter. The atoms are all in
the ground state before the pulse and assuming a perfect $\pi$ pulse
will end up in the excited state afterward regardless. Any small
coherent excitation caused by imperfect $\pi$ pulses can be ignored,
it will quickly dephase because of the inhomogeneous broadening and
will not be rephased as an echo  until after the second $\pi$ pulse which
is outside the region of time of interest. The ASE caused by the
inversion due to this $\pi$-pulse will be spatially multimode, with
the amount of ASE in a particular mode determined by the gain
experienced traversing the sample. 

Suppose we set a detection system to look at the ASE produced with
wavevector $\mathbf{k}_\text{ASE}$, from Eqs.~\ref{equ:ex1parax}
and~\ref{equ:ex2parax}. We can see that the light with this wavevector
is entangled with the atomic excitation with mode
$-\mathbf{k}_\text{ASE}$. The $\pi$-pulse transfers this to the
wavevector $-\mathbf{k}_\text{ASE}+2\mathbf{k}_\pi$ according to
Eq.~\ref{equ:paraxpi}. Equations~\ref{equ:gs1parax_diag}
and~\ref{equ:gs2parax_diag} connect atomic and optical modes with the
same wavevector so we have that the wavevector for the RASE is $\mathbf{k}_\text{RASE} =
-\mathbf{k}_\text{ASE}+ 2\mathbf{k}_\pi$ or 
\begin{equation}
  \label{eq:phasematch}
  \mathbf{k}_\text{ASE} + \mathbf{k}_\text{RASE} =  2\mathbf{k}_\pi.
\end{equation}

This is the same phase matching condition as a two pulse photon echo,
$\mathbf{k}_\text{input} + \mathbf{k}_\text{echo} =
2\mathbf{k}_\pi$. While this phase matching condition is valid outside
the paraxial regime, the only way to achieve phasematching is with the
beams colinear or close to conlinear, because the ASE the RASE and the $\pi$ pulse must all be at the same
frequency.

\subsection{\bf DLCZ }

\begin{figure}
  \includegraphics[width=0.45\textwidth]{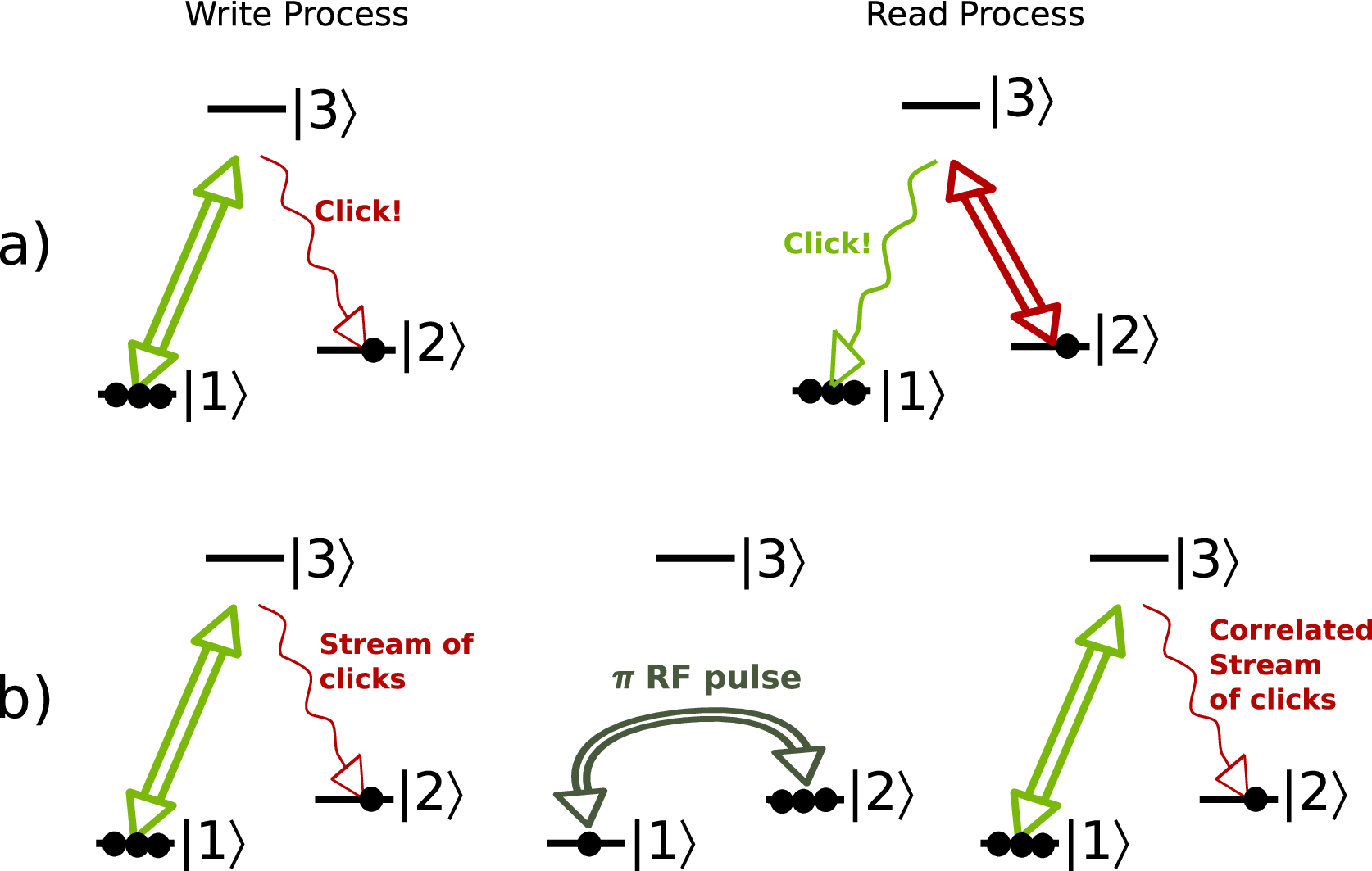}
  \caption{(Color online) a) DLCZ protocol showing write and read process. b) Modified protocol. The inhomogeneous broadening of the $|1\rangle$-$|2\rangle$ transition now leads to an increase in bandwidth. \label{fig:dlcz_new}}
\end{figure}

It is interesting to consider the relationship of the current scheme with the DLCZ protocol \cite{dlcz}. The DLCZ protocol involves the creation of entanglement between distant ensembles. The relevant energy level diagrams are shown in Fig.~\ref{fig:dlcz_new}a. Once the level $|3\rangle$ has been adiabatically eliminated the write process is formally equivalent to a set of excited state atoms ($|1\rangle$) spontaneously emitting into the level $|2\rangle$. The emitted optical field is then steered elsewhere for entanglement generation with another ensemble of atoms \cite{dlcz}. Once entanglement is generated between two ensembles, one wishes to read out one ensembles atomic field to a photonic field in order to implement entanglement swapping \cite{dlcz}. For the read process, state $|2\rangle$ becomes the excited state and state $|1\rangle$ the ground state.

One problem with this is the inhomogeneous broadening of the
$|1\rangle$-$|2\rangle$ transition causes dephasing limiting the time
separation between the writing and reading process. A modified DLCZ
protocol, in close analogy with RASE, would overcome this problem. A
rephasing pulse on the $|1\rangle$-$|2\rangle$ transition utilizes the
inhomogeneous broadening, now increasing the bandwidth of the process
rather than reducing the time separations. The sequence of events for
this modified DLCZ protocol are shown in Fig.~\ref{fig:dlcz_new}b. It
is worth noting that the modified DLCZ protocol does not have the same
issue with echo efficiency as the two level scheme because the
classical coupling field can be altered meaning that the ensemble can
be optically thin for the writing process and thicker for the reading
process.  

The phase matching conditions for the modified DLCZ protocol
will be the same as given in  Eq.~\ref{eq:phasematch} for RASE. 
However with a Raman transition it is the wavevector difference for the two
optical fields that is important. This means that one has alot more
freedom in the implementation because one isn't restricted by the
requirement that $\omega = ck$, as one is in the two level case.

\section{\bf Conclusion}

In conclusion we have shown that rephased amplified spontaneous
emission has strongly non-classical correlations with the original
amplified spontaneous emission in the optically thin regime. This
leads to the possibility of a modified DLCZ protocol, where the
problem of dephasing due to inhomogeneous broadening of the hyperfine
transitions is solved by a rephasing pulse, increasing the bandwidth
of the process.



\end{document}